\documentclass[11pt]{article}
\usepackage{authblk}
\usepackage{graphicx}
\usepackage{xcolor}
\usepackage[margin=1in]{geometry}
\usepackage{natbib}
\usepackage[hidelinks]{hyperref}

\usepackage{wrapfig}
\usepackage{comment}
\definecolor{esoblue}{RGB}{0,62,126}

\title{\textbf{Spectroscopic Alerts for the Time-Domain Era}}

\author[1]{Alejandra Melo (\texttt{alejandra.melo@eso.org})}
\author[1]{Paula Sanchez-Saez}
\author[1]{Valentin D. Ivanov}
\author[2]{Richard I. Anderson}
\author[1]{Amelia Bayo}
\author[2]{Avraham Binnenfeld}
\author[3]{Sofia Bisero}
\author[4,5]{Dragana Ili\'c}
\author[4]{Andjelka B. Kova\v cevi\'c}
\author[6]{Fatemeh Zahra Majidi}
\author[7,8]{Jaroslav Merc}
\author[1]{Anna Pala}
\author[9]{Swayamtrupta Panda}
\author[10]{Sarath Satheesh-Sheeba}
\author[3]{Fabian Sch\"ussler}
\author[11]{Susanna D. Vergani}

\affil[1]{European Southern Observatory, Karl-Schwarzschild-Strasse 2, 85748 Garching bei München, Germany}
\affil[2]{Institute of Physics, \'Ecole Polytechnique F\'ed\'erale de Lausanne (EPFL), Observatoire de Sauverny, 1290 Versoix, Switzerland}
\affil[3]{IRFU, CEA, Université Paris-Saclay, France}
\affil[4]{Department of Astronomy, Faculty of Mathematics, University of Belgrade, Studentski trg 16, 11000 Belgrade, Serbia}
\affil[5]{Hamburger Sternwarte, Universit\"at Hamburg, Gojenbergsweg 112, D-21029 Hamburg, Germany}
\affil[6]{INAF OACN, Italy}
\affil[7]{Astronomical Institute of Charles University, Czech Republic}
\affil[8]{Instituto de Astrofísica de Canarias, Tenerife, Spain}
\affil[9]{International Gemini Observatory, NSF NOIRLab, Casilla 603, La Serena, Chile}
\affil[10]{Instituto de Astrofísica, Facultad de Ciencias Exactas, Universidad Andres Bello, Fernández Concha 700, 7591538 Las Condes, Santiago, Chile}
\affil[11]{Observatoire de Paris, Université PSL, CNRS, Sorbonne Université, Meudon, 92190, France}

\date{December 2025}

\begin{document}

\maketitle

\begin{abstract}
Time-domain astronomy is entering an era of unprecedented discovery driven by wide-field, high-cadence surveys such as LSST, Roman, \textit{Euclid}, SKA, and PLATO. While some of these facilities will generate enormous photometric alert streams, the physical interpretation of variability and transients often requires spectroscopy, which encodes changes in ionisation state, kinematics, and accretion that are inaccessible to photometry alone. A critical gap is therefore emerging: next-generation surveys may produce up to $\sim10^9$ alerts per year, whereas global spectroscopic follow-up is limited to only $\sim10^4$--$10^5$ transient spectra annually. We present the concept of \emph{spectroscopic alerts}: real-time notifications triggered by significant spectral evolution, enabling spectroscopy to act as a discovery channel rather than solely as follow-up. We outline the key science cases enabled by this capability and describe the instrumental and operational requirements of a wide-field, highly multiplexed spectroscopic facility capable of delivering real-time spectral discovery for 2040s time-domain and multi-messenger astronomy.

\end{abstract}

\thispagestyle{empty}
\newpage
\setcounter{page}{1}

\newpage

Time-domain astronomy is entering a new era. Over the next two decades, major survey facilities, such as the Vera C. Rubin Observatory (LSST, \citealt{2019LSST}), the Nancy Grace Roman Space Telescope (Roman, \citealt{2015Roman}), \textit{Euclid} (\citealt{2025Euclid}), the Square Kilometre Array  \citep{2009SKA}, the PLAnetary Transits and Oscillations of stars (PLATO, \citealt{2025PLATO}), and WEAVE \citep{2024Jin} will deliver continuous, multi-wavelength monitoring of the sky with unprecedented depth, cadence, and coverage, building on the legacy of space-based time-domain missions that have demonstrated the power of long, homogeneous, high-precision observations across diverse astrophysical phenomena \citep{2025Huber}. During their operational lifetimes, some of these facilities will generate a massive alert avalanche detecting flux variations and moving objects, while their survey depth, uniformity, and multi-wavelength coverage will remain a foundational resource for time-domain studies.

Interpreting this rapidly evolving sky requires access to the physical information encoded in spectral evolution. Such changes often reflect orders-of-magnitude shifts in physical conditions that cannot be constrained by photometry alone, for example in accretion rates, variability in young stellar and planetary mass objects, or the changing-look phases of Active Galactic Nuclei (AGNs). LSST broadband filters detect only the most extreme cases, while most of the underlying physics remains hidden without spectroscopy. 

\begin{wrapfigure}{r}{0.63\textwidth}
    \centering
    \vspace{-10pt}
    \hspace*{-1.0cm}
    \includegraphics[width=0.60\textwidth]{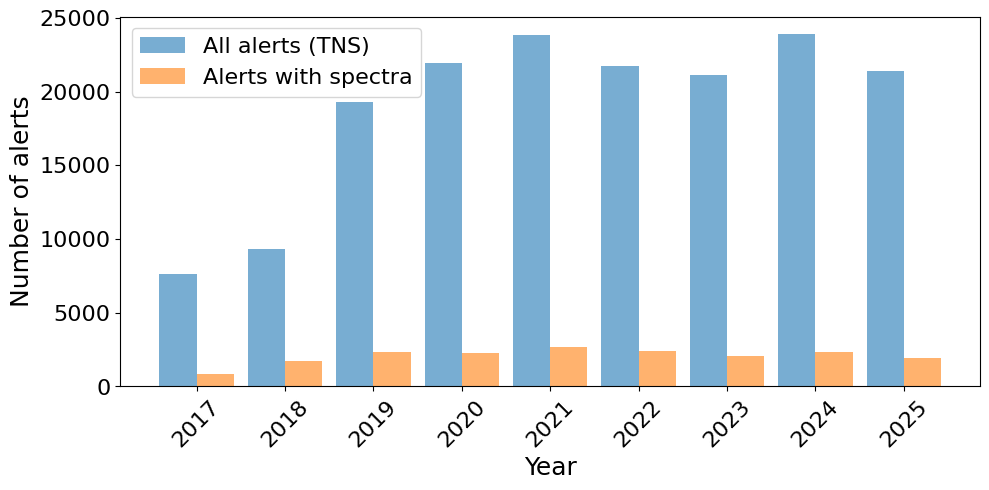}
    \vspace{-8pt}
    \caption{\small
        Comparison of all Transient Name Server (TNS) alerts and the subset with spectra (2017–2025). The large gap between alerts and spectroscopic classifications highlights the significant global shortfall in follow-up capacity.
    \vspace{-10pt}
    }
    \label{fig:tns_alerts_spectra}
\end{wrapfigure}

The expanding network of multi-messenger and multi-wavelength facilities, including upcoming Gravitational Waves (GW) observatories such as the Laser Interferometer Space Antenna (LISA, \citealt{2017LISA}), will further increase the volume and diversity of transient detections. However, it also exposes a critical gap: while imaging surveys provide large-scale photometric alert streams, there is no equivalent mechanism for spectral discovery. Existing and planned infrastructures lack the means to obtain \textbf{timely, systematic, high-cadence spectroscopy}. Next-generation surveys are expected to generate up to $\sim$10$^9$ alerts per year \citep{2019LSST}, whereas global spectroscopic capacity is limited to only $\sim$10$^4$–10$^5$ transient spectra annually \citep{2019Nordin,2020Fremling}. This severe mismatch (Fig. \ref{fig:tns_alerts_spectra}) highlights the need for spectroscopic capabilities designed to meet the demands of the emerging discovery landscape.

Existing spectroscopic facilities, with limits in multiplexing, field of view, and operational flexibility, cannot deliver the rapid, wide-field spectral characterization required in this new regime. More importantly, no facility currently provides \textbf{spectroscopic alerts}: real-time notifications that a source has undergone a significant spectral change, such as the appearance of new emission/absorption components, rapid ionisation-state transitions, or evolving velocity structure. Detecting them, as they occur, would turn a spectroscopic facility into a genuine discovery instrument, opening a pathway to fundamentally new insights into the evolving Universe.

Existing surveys such as SDSS \citep{2000SDSS}, LAMOST \citep[][in their southernmost fields]{2012RAA....12.1197C}, DESI \citep{2016DESI}, 4MOST \citep{20144MOST}, PFS \citep{2016tamura}, and even multi-epoch efforts like SDSS-V \citep{2025Kollmeier} excel at large predefined samples but lack the cadence, flexibility, and autonomous response needed for real-time spectral monitoring or generating spectroscopic alerts. Dedicated time-domain efforts, such as the 4MOST Time-Domain Extragalactic Survey (TiDES; \citealt{2025TiDES}), demonstrate the scientific value of systematic spectroscopic monitoring, but their limited scale, only a few thousand transients and AGNs per year, highlights the need for a facility capable of delivering spectroscopic alerts at LSST-level volumes.

In addition, many high value transients from GW, neutrinos, or high-energy facilities also demand integral field spectroscopic (IFS) capabilities to identify counterparts within large localisation regions. A high-multiplexing facility such as the proposed Wide-field Spectroscopic Telescope (WST; \citealt{2024WST}), with streamlined operations, rapid data processing, and very high fiber multiplexing and IFS capabilities, would enable nightly acquisition of vast numbers of spectra and the automatic detection and distribution of alerts driven by spectral evolution, delivering sustained high-cadence spectroscopy for tens to hundreds of thousands of dynamically evolving sources.

\vspace{0.2cm}

{\large\bfseries \textcolor{esoblue}{Science cases enabled by real-time spectroscopic alerts}}\par\vspace{0.2cm}

\begin{figure}[htb!]
    \centering
\includegraphics[width=0.51\linewidth]{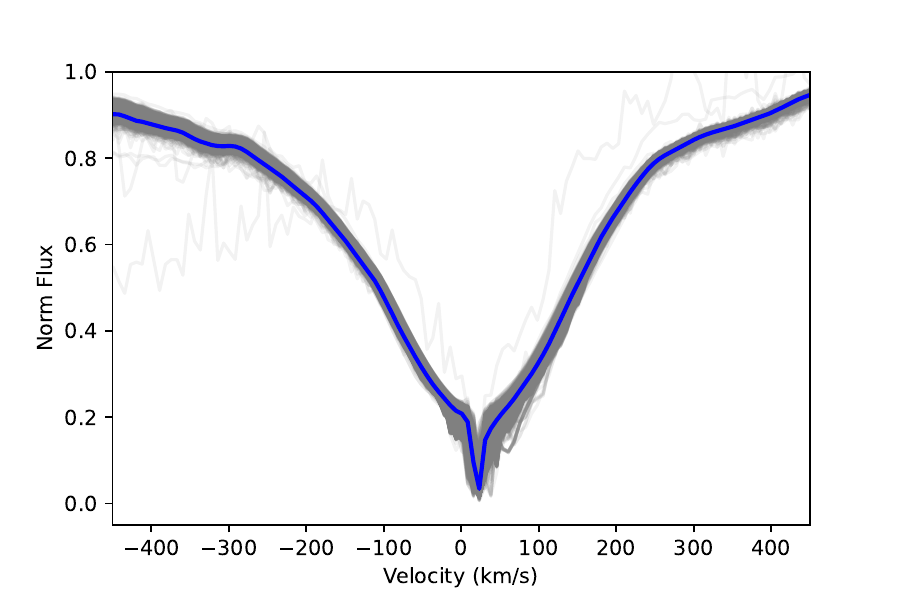}
\includegraphics[width=0.48\linewidth]{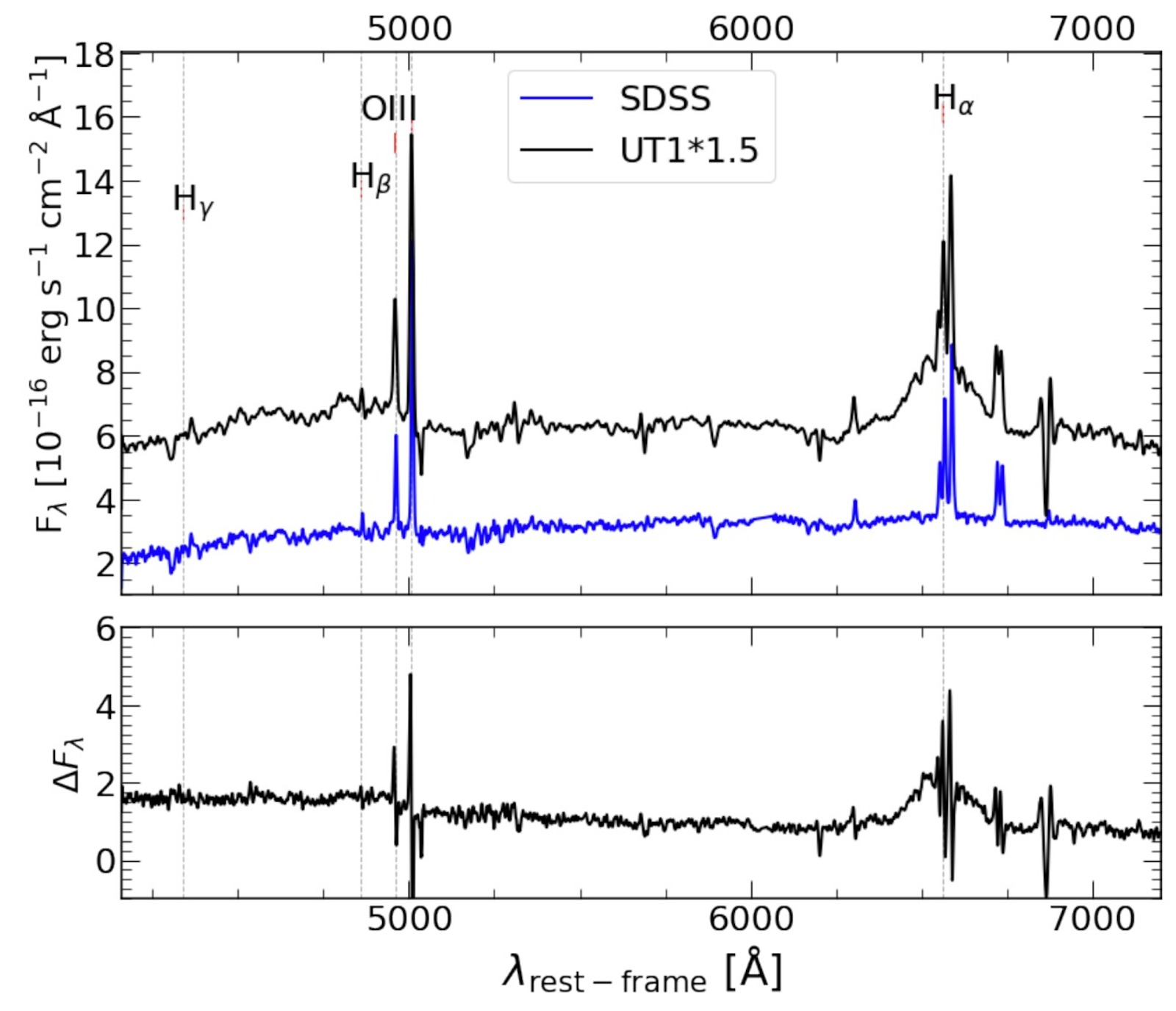}
\caption{\small 
Examples of pronounced and evolving spectral variability.
\textit{Left:} Simulation showing the detectability of the Beta Pictoris' exocomets with the expected WST high resolution mode. Individual epochs are shown as gray lines; additional redshifted and some blue shifted components appear and disappear on time scales of hours to days (WST collaboration, private communication). 
\textit{Right:} Changing-look AGN candidates exhibiting changes in Balmer and [O{\sc iii}] emission-line strengths between SDSS and follow-up spectra (Adapted from \citealt{2022Lopez-Navas}).}
    \label{fig:examples}
\end{figure}

In the stellar regime, evolving line profiles can trace pulsations, accretion bursts, exocomet transits, circumstellar gas flows, winds, and short-lived obscuration and flare events on hour–day timescales. As illustrated in Fig.~\ref{fig:examples} (left), high-cadence, high-resolution spectroscopy enables the detection of short-timescale spectral variability in planetary systems, revealing activity from infalling exocomets that is inaccessible to photometry alone and providing a unique probe of planetary system evolution.

Explosive and nuclear transients similarly require timely spectroscopy to unlock their physical interpretation. The earliest phases of supernovae, tidal disruption events, and fast transients exhibit rapidly evolving ionisation states, velocities, and continuum shapes that reveal the structure of the progenitor, the surrounding material, and the physics driving the explosion. In AGNs and other nuclear transients, real-time spectral monitoring enables the detection of changing-look events and Ansky-type sources \citep{2024SanchezSaez}, as well as broad-line region variability, directly tracing accretion-state transitions and changes in black hole environments (Fig.~\ref{fig:examples}, right). Long-term spectroscopic monitoring can further reveal the presence of close binary supermassive black holes and provide insight into black hole growth and feedback.

Identifying electromagnetic counterparts to gravitational-wave events is challenging due to the large localisation regions, which may contain thousands of galaxies and photometrically variable sources. Highly multiplexed spectroscopy enables efficient, galaxy-targeted searches within these large error volumes, effectively turning wide-field facilities into discovery instruments and providing the only practical way to identify fast-fading counterparts for detailed characterisation rapidly.

Finally, time-resolved spectroscopy enables the detection of phenomena that fall outside currently recognised classes of variability. Automated identification of unexpected spectral behaviour will uncover rare events and may reveal entirely new types of astrophysical activity, for example, potential signatures of close binary supermassive black holes, which have not yet been definitively confirmed. A spectroscopic alert system, therefore, supports not only established science goals but also the discovery of phenomena that cannot yet be predicted.

\vspace{0.2cm}

{\large\bfseries \textcolor{esoblue}{A spectroscopic alert system as a key capability for 2040s time-domain science}}\par\vspace{0.2cm}

Meeting these scientific goals requires more than high-cadence spectroscopy: it demands an integrated system that can identify and interpret spectral changes in real time and place them in their broader astrophysical context. This entails rapid data processing combined with automated analysis, and the ability to link new spectra to multi-epoch and multi-wavelength information from surveys such as \textit{Gaia}, LSST, \textit{Roman}, 4MOST, DESI, SDSS, and LAMOST, as well as to alerts from emerging multi-messenger facilities including next-generation gravitational-wave, neutrino, and high-energy observatories (e.g., Einstein Telescope, Cosmic Explorer, LISA, CTAO, SKAO, KM3NeT, and IceCube).

Delivering these capabilities at the scale required by 2040s time-domain surveys demands a wide-field, highly multiplexed spectroscopic facility. A WST-like facility, with a 10-12m class wide-field telescope equipped with $\sim$30,000 fibers and spectral resolution $R\sim3,000–10,000$ would provide the cadence, volume, and operational flexibility needed to capture spectral evolution across a wide range of timescales, from hour-level changes in kilonovae, early supernovae, accretion bursts, and exocomet transits to the daily–weekly variability of AGN and stellar systems.

Together, these elements define the core requirements for the next-generation spectroscopic alert ecosystem: rapid availability of spectral data, contextualised interpretation, meaningful and reliable classification, synergy with other facilities, and the ability to follow variability across an exceptionally wide dynamic range of timescales. This framework would fundamentally reshape how the community engages with an evolving sky, enabling fast, coordinated, and scientifically optimised responses to the most informative events, transforming this new rich data stream into physical understanding.

More broadly, spectroscopic alerts have strategic implications for ESO’s archive and data-distribution paradigm. Enabling alerts driven by spectral evolution requires archives to support persistent time-domain spectroscopy, cross-epoch linking, and access to evolving spectral information, rather than treating spectra as isolated end products. Embedding these capabilities within ESO’s infrastructure, relevant to 4MOST and future spectroscopic facilities alike, would position ESO’s archive as an active interface between observations and discovery, and as a cornerstone of time-domain and multi-messenger astronomy in the coming decades.

\vspace{+0.3cm}

{\large\bfseries \textcolor{esoblue}{Acknowledgements}}\par\vspace{0.2cm}

This project has received funding from the European Union Horizon Europe Research and Innovation Action under grant agreement no. 101183153-WST. Views and opinions expressed are however those of the author(s) only and do not necessarily reflect those of the European Union or the European Research Executive Agency (REA). Neither the European Union nor the REA can be held responsible for them. RIA is funded by the Swiss National Science Foundation through an Eccellenza Professorial Fellowship (award PCEFP2$\_$194638). S.B. acknowledges the Astrophysics Centre for Multi-messenger studies in Europe (ACME), funded under the European Union’s Horizon Europe Research and Innovation Programme, under grant agreement No. 101131928. D.I and A.B.K. acknowledge funding provided by the University of Belgrade - Faculty of Mathematics through the grant (the contract 451-03-136/2025-03/200104) of the Ministry of Science, Technological Development and Innovation of the Republic of Serbia. JaM was supported by the Czech Science Foundation (GACR) project no. 24-10608O. SP is supported by the international Gemini Observatory, a program of NSF NOIRLab, which is managed by the Association of Universities for Research in Astronomy (AURA) under a cooperative agreement with the U.S. National Science Foundation, on behalf of the Gemini partnership of Argentina, Brazil, Canada, Chile, the Republic of Korea, and the United States of America. FS acknowledges the support of the French Agence Nationale de la Recherche (ANR) under reference ANR-22-CE31-0012.

\bibliography{references}

\end{document}